\documentclass[12pt,preprint]{aastex}

\shorttitle{Bar-Halo Interaction and Bar Growth}
\shortauthors{Athanassoula}

\begin{document}


\title{Bar-Halo Interaction and Bar Growth}


\author{E. Athanassoula}
\affil{Observatoire de Marseille, 2 Place Le Verrier, 13248 Marseille
cedex 04, France}



\begin{abstract}
I show that strong bars can grow in galactic discs, even when the
latter are immersed in haloes whose mass
within the disc radius is comparable to, or larger than, the mass of the
disc. I argue that this is due to the response of
the halo and in particular to the destabilising influence of the halo
resonant stars. Via this instability mechanism the halo can stimulate,
rather than restrain, the growth of the bar. 
\end{abstract}


\keywords{galaxies: evolution, galaxies: halos, galaxies: kinematics and dynamics}


\section{Introduction}

Galactic discs are generally unstable and form bars within a few
dynamical timescales (e.g. Athanassoula 1984, or Sellwood \&
Wilkinson 1993, for reviews). The first remedy proposed against this
instability is to immerse the disc in a massive halo component.
Ostriker \& Peebles (1973), from an insightful
analysis of $N$-body simulations with a very small number of
particles, were the first to outline the 
stabilising influence of haloes. They further suggested that halo-to-disc
mass ratios of 1 to 2.5 (interior to the disc) are required for stability. 
Athanassoula \& Sellwood (1986) measured the growth rates of the most
unstable bar-forming modes in 2D $N$-body simulations and
showed that the growth rate is 
smaller in cases with large halo-to-disc mass ratios. Unfortunately,
no such analysis has been made for fully self-consistent 3D $N$-body
simulations. 
On the other hand, Athanassoula \& Misiriotis (2002, hereafter AM) and
Athanassoula (2002a,b) argue that stronger bars can grow
in simulations with initially very high halo-to-disc mass ratios than
in simulations with lower halo-to-disc mass ratios.
In order to understand whether the above pieces of evidence are 
contradictory, and, if so, to attain a coherent picture,
I will examine in this letter the role of the halo in the formation of
the bar. Several papers have focused on the influence of the
halo on the slowing down of the bar (e.g. Tremaine \& Weinberg 1984,
Weinberg 1985, Combes et al. 1990, Little \& Carlberg 1991, Hernquist
\& Weinberg 1992, Athanassoula 1996, Debattista \& Sellwood 1998 and
2000). This paper is complementary, since it concentrates on the role
of the halo in determining the bar growth. 
The discussion will focus on three simulations. The results, however,
are based on a much larger set of simulations, leading to the same
conclusions.   

\section{Results on bar formation}
\label{sec:simul}

I will describe the results of three fully self-consistent 3D $N$-body
simulations 
of isolated galaxies consisting of a disc and a halo component. The initial
conditions and the numerical methods are described in AM,
whose notation I follow here. The
three initial conditions have identical exponential 
discs of unit mass and scale length, scale height $z_0$ = 0.2 and $Q$ =
1. Their halos are five 
times as massive as their discs and have different degrees of central
concentration. They are initially isotropic, spherical and
non-rotating, with a radial density profile

\begin{equation}
\rho_h (r) = \frac {M_h}{2\pi^{3/2}}~~ \frac{\alpha}{r_c} ~~\frac {exp(-r^2/r_c^2)}{r^2+\gamma^2}.
\end{equation}

\noindent
with $M_h$ = 5,  $r_c$ = 10. The parameter $\alpha$ is a normalisation
constant defined by 

\begin{equation}
\alpha = [1 - \sqrt \pi~~exp (q^2)~~(1 -erf (q))]^{-1},
\end{equation}

\noindent
where $q=\gamma / r_c$ and $erf$ is the error function (Hernquist
1993). The initial circular 
velocity curves of the three models are shown in the upper 
panels of Fig.~1. The first simulation, hereafter MD, has $\gamma$ =
5, and its disc dominates in the inner parts, roughly up to $r$ = 5. In
the other two (hereafter 
MH and RH, respectively), $\gamma$ = 0.5, and the halo contribution is
slightly larger  
than that of the disc up to the maximum of the disc rotation curve, and
considerably so at larger radii.
Model RH is identical to model MH,
except that its halo is rigid -- i.e. given by a potential imposed
on the disc particles -- and does not evolve during the simulation.
The number of particles in the simulations is 1131206, 1163030
and 200000, respectively. The simulations were run on a GRAPE-5 system
with a softening of 0.0625 and a time-step  of 0.015625. I assessed
the numerical robustness of my results by trying
double the number of particles, different values of the softening
and time step, as well as direct summation and a non-GRAPE tree code.
For a reasonable calibration (AM),  
$t$ = 500 corresponds to 7 $\times 10^9$ years.  
Different values, however, can be obtained for different scalings of 
the disc mass and scale length.

The second row of panels of Fig.~1 gives the three circular
velocity curves at time $t$ = 800. Note that the disc material has moved
considerably inwards in the two first cases, but very little
in the third one. In model MD the difference between times 0 and 800 is
quantitative, since the disc dominates the inner parts both initially and
after the evolution, albeit to a larger degree after the bar has grown. On 
the other hand for model MH the difference 
is qualitative, since at $t$ = 800 the inner regions are dominated by
the disc, contrary to what was the case initially. The third row of
panels gives the face-on view of the disc component. The differences
between the properties of the three models are striking. A
comparison of models MD and MH shows that {\it the bar that grew
  in the initially more halo-dominated environment is stronger 
  than the bar that
  grew in the disc-dominated environment}. It is longer and its
isophotes are more rectangular-like. The strength of the bar can be
measured with the help of the relative amplitude of the
Fourier components of the density or mass (AM), and the last row of
panels in Fig.~1 shows that indeed the MH bar is considerably stronger
than the MD one. The $m$ = 4, 6 and 8  
Fourier components of the face-on density of model MH are
also considerably stronger than those of model MD (AM). 

Very strong differences are also found when comparing models MH
and RH. Model MH has a very strong bar, while model RH has a
very mild oval in the innermost parts. Their edge-on views are also very
different (not shown here, cf. Athanassoula 2002b for another
example). Model MH seen side-on has a strong peanut or `X'-shape, and
a big bulge-like protuberance if seen end-on. Model RH shows no such
features. Finally the difference between their integrated Fourier
components is striking. 

The three models differ also in the way their bars evolve. In model MD the bar
grows very rapidly during the first part of the evolution, but little,
if at all, after that. On the contrary the growth of the bar in model MH 
during the first part of the evolution is slower than in MD, but stays
considerable, although less important than in the first part, till the
end of the simulation.  The pattern
speed of model MH starts off higher than that of model MD, but ends
smaller. The pattern speed of MD also decreases with
time, but less so (cf. Debattista \& Sellwood 1998, 2000). The pattern 
speed of the non-axisymmetric component in  
RH can not be measured reliably before $t$ = 300, and after 
that does not show any signs of decrease. For models MD and MH
there is exchange of energy and
angular momentum between the disc and halo components, so that the
halo, which was initially non-rotating, displays rotation after the
bar has grown. This is small for model MD and considerable for model
MH, for which at $t$ = 900 the halo has somewhat less than
a half of the angular momentum of the disc, i.e. not far from a
third of the total. 
 
\section{The role of the halo}
\label{sec:halo}

In order to understand the role of the halo in the formation and
evolution of the bar, I froze the
potential at four selected times during each simulation and chose randomly
100\,000 disc and 100\,000 halo particles. I followed their orbits during
40 bar rotation periods and
calculated their basic frequencies, namely
the angular frequency $\Omega$, the epicyclic frequency $\kappa$ and
the vertical frequency $\kappa_z$. For the two latter ones I used a
spectral analysis technique (Binney \& Spergel 1982, Laskar
1990). In most cases there were several secondary peaks, and the
frequency was determined by the main one. The angular
frequency proved more difficult to calculate 
reliably, so I supplemented 
the spectral analysis with other, more straightforward methods, based
on following the angle as a function of time. Agreement between
the values for the angular frequency obtained by the various methods
was found to be satisfactory for 85\% or more of the particles.

An orbit is resonant if there are three integers
$l$, $m$ and $n$ such that 
$l \kappa + m \Omega + n \kappa_z = m \Omega_p$, where $\Omega_p$ is 
the pattern speed of the bar. I will here
restrict myself to radial (planar) resonances, for which $n$ = 0. The most
important such resonances are the inner
Lindblad resonance (ILR), where $l$ = -1 and $m$ = 2, corotation
resonance (CR), where $l$ = 0, and the outer Lindblad resonance (OLR),
where $l$ = 1 and $m$ = 2. At CR the
particles have the same angular frequency as the bar, while at the
other resonances they make $m$ radial oscillations in the time they
make $l$ revolutions around the center of the galaxy. 
Fig.~2 shows the number of particles (orbits),
$N_R$, that have a frequency ratio $R = (\Omega - \Omega_p) / \kappa$
within a bin of a given width centered on a value of this ratio, 
plotted as a function of $R$.

Let me first describe the results for the disc components of MD and MH. 
The distribution in both cases is far from homogeneous, with 
strong peaks at the location of the main resonances. The highest peak,
both for the MH and the MD disc, is for $(\Omega - \Omega_p) / \kappa$
= 0.5, i.e. at the ILR. Indeed, orbits
making two radial oscillations in the time they make one revolution
around the center of the galaxy are the backbone of the bar. In
principle $x_2$ type orbits (cf. Contopoulos \& Grosb{\o}l 1989 for a review)
could also be found in this peak. We have, however, verified that the
vast majority of the orbits here are $x_1$ type. The ILR peak
in the MH case is roughly 1.5 times higher than in MD, consistent with
the fact that the bar in the MH case is stronger.
Model MD has a sizeable CR peak and a much
lower one at the OLR, while model MH has only a small CR peak and no
OLR one. The ratio of the height of the CR peak to that
of the ILR peak is roughly 0.6 for the MD case, while for model MH it is only
0.09. These differences are, to a large extent,
due to the differences in the corotation radii of the two models. 
Thus at time 800 the CR radius is 7.1 for model
MH and 4.6 for model MD. Similar differences are found for the
OLR radii of the two models. Therefore these two resonances are in the
outer parts of the MH disc and can trap only few particles. This
is not the case for model MD, and the differences in the trappings are
reflected in the differences in the heights of the respective resonant peaks. 
Model MH has also a clear peak at the inner 1:3 resonance and
perhaps one at the 1:4. Model MH has a clear inner 1:4 peak.

The big surprise, however, comes from the halo component. So far
considered as non-, or little, responsive, it shows, on the contrary,
unmistakable signs of strong resonances with the bar. For both models
the strongest resonance is CR, which proves to be much stronger in the
halo than in the disc component. In fact the CR peak in the halo
of model MH is higher than any of the disc peaks in MD. There
are also sizeable peaks at the OLR. 

There are important differences between the orbital structures of the MH and MD
haloes, as was the case for the corresponding discs. The ILR peak of
model MH is relatively high -- the second highest peak for this model
and time -- and, in
general, there is considerably more material between 
ILR and CR than in model MD, while the CR peak is 40\% higher than in
model MD. All these are in agreement with the fact
that the MH halo is much more concentrated than the MD one, while 
the MH corotation is at a larger radius. For smaller values of $R$
the situation shifts and the peaks are stronger for the MD
case. The OLR peak of model MD is more than 1.8 times 
as high as the MH one, and the -1:1 peak is also clear in MD,
while for model MH this resonance and its surroundings have been
depleted. 

The relative and absolute heights of the resonance peaks
vary with time. However, a large fraction of the orbits which are
within a resonance peak at a given time continue to be so at a later
time. Thus, comparing times 500 and 560, I find that 60 to 80 per cent of
all orbits in one of the main resonant peaks at time 500 are found in the
same peak at time 560. Similar numbers can be found by comparing times
800 and 860, except for the CR peak of the halo in the MH
simulation, where the fraction falls to 55 per cent. Seen the
uncertainties in delimiting the extent of a frequency peak, and the
uncertainties in the calculation of the frequencies, these numbers are
compatible with a large trapping of particles in the resonances. A
more detailed description of these changes, together 
with a discussion on their effect on the structure and the dynamical
evolution of the galaxy will be given elsewhere.

\section{Discussion}
\label{sec:discuss}

In the above I compared three simulations starting off with
identical discs, but different halo components. Any differences in
their dynamical evolution should thus be attributed to the haloes.
The strongest bar forms in the most halo-dominated case, provided this
is live, while the simulation with an identical but rigid halo forms
only a mild oval distortion, in the inner parts only. This very big
difference can be attributed only to the responsiveness of the halo. 
The disc-dominated case, i.e. the case with the less important halo
component, formed an intermediate bar. I thus reach the interesting 
conclusion that haloes can, at least in some cases, stimulate the bar
instability and lead to stronger bars. 

Lynden-Bell \& Kalnajs (1972) have shown that disc stars at resonances
can absorb or emit angular momentum, thus driving the dynamical
evolution. Halo stars in resonance can also exchange
energy and angular momentum (Tremaine \& Weinberg 1984). In general,
the halo resonances will absorb angular momentum. Prompted by such
considerations I searched for resonant stars in the halo, and found 
large numbers (at least after the bar has grown). Since these can
exchange energy 
and angular momentum with disc resonant stars, they can stimulate the
bar instability, contrary to previous beliefs, and thus explain why
stronger bars can grow in more halo-dominated surroundings. The
situation is particularly clear in the case of model MH, where, as
Fig.~2 shows, there are hardly any absorbers in the disc to take the
angular momentum emitted by the stars at the ILR. This task is thus
necessarily performed by the halo resonant stars. 

Since the bar is inside corotation, it has negative energy and angular
momentum (e.g. Lynden-Bell \& Kalnajs 1972), and if it emits angular
momentum it will be destabilised, i.e. it will in general grow stronger. This
is in good agreement with the results on angular momentum transfer
discussed at the end of \S2. These show that the halo takes
positive angular momentum from the disc/bar component and, since the
disc, the bar and the final halo component all rotate in the same direction, 
this will destabilise the bar.

In the initial phases of the evolution the bar grows faster
in the disc dominated surroundings, in good agreement with previous
results (e.g. Athanassoula \& Sellwood 1986). However, in fine, the
bar in the MH model reaches a higher amplitude, due to the stronger 
exchange of energy and angular momentum with the resonant halo stars. 
Thus a massive halo may help the bar grow and
become very strong, so that very strong bars may be found in initially
halo-dominated galaxies. In other words there is no disagreement
between the results of e.g. Athanassoula \& Sellwood (1986) and those
presented here or in AM or in Athanassoula (2002a, b).

Hernquist \& Weinberg (1992) checked the angular momentum given to a
live halo by a rigid bar turned first on and then off adiabatically. In
spite of considerable noise, they found indications that the angular 
momentum was deposited primarily at the resonances and that it is
mainly absorbed. This tentative result is clearly established
here, where both the disc and the halo are live, the
evolution is not artificially forced and the resonances are clearly
outlined. 

In the linear theory the exchange of energy and angular momentum is
linked with the growth of the wave. In the nonlinear theory on
the other hand it has been linked to the slowdown of the bar
(Tremaine \& Weinberg 1984, Weinberg 1985, Hernquist \& Weinberg 1992). 
Both effects are clearly visible in the simulations presented here and
should be physically linked. 

The orbital structures revealed in Fig.~2 are not specific to
the models discussed above. I have repeated a similar analysis for
other times and other simulations and found similar behaviours. 
They should thus be representative of a wide class of models. 

I am thus proposing a new instability mechanism, by which the
halo will stimulate, rather than restrain, bar growth. This of course
will only work 
if the halo is non-rigid and is capable of absorbing positive angular
momentum. Similarly the bar should also be non-rigid. Further analysis
of this mechanism will be given elsewhere.

\acknowledgments

I would like to thank A. Bosma, M. Tagger and F. Masset for
stimulating discussions, and  A. Misiriotis and J.~C. Lambert for
their help with the software calculating the orbital frequencies. I
would also like to thank the Region PACA, the
INSU/CNRS, the University of Aix-Marseille I and the IGRAP for funds to develop
the GRAPE computing facilities used for the simulations discussed in
this paper.

\begin{figure} 
\epsscale{0.85}
\plotone{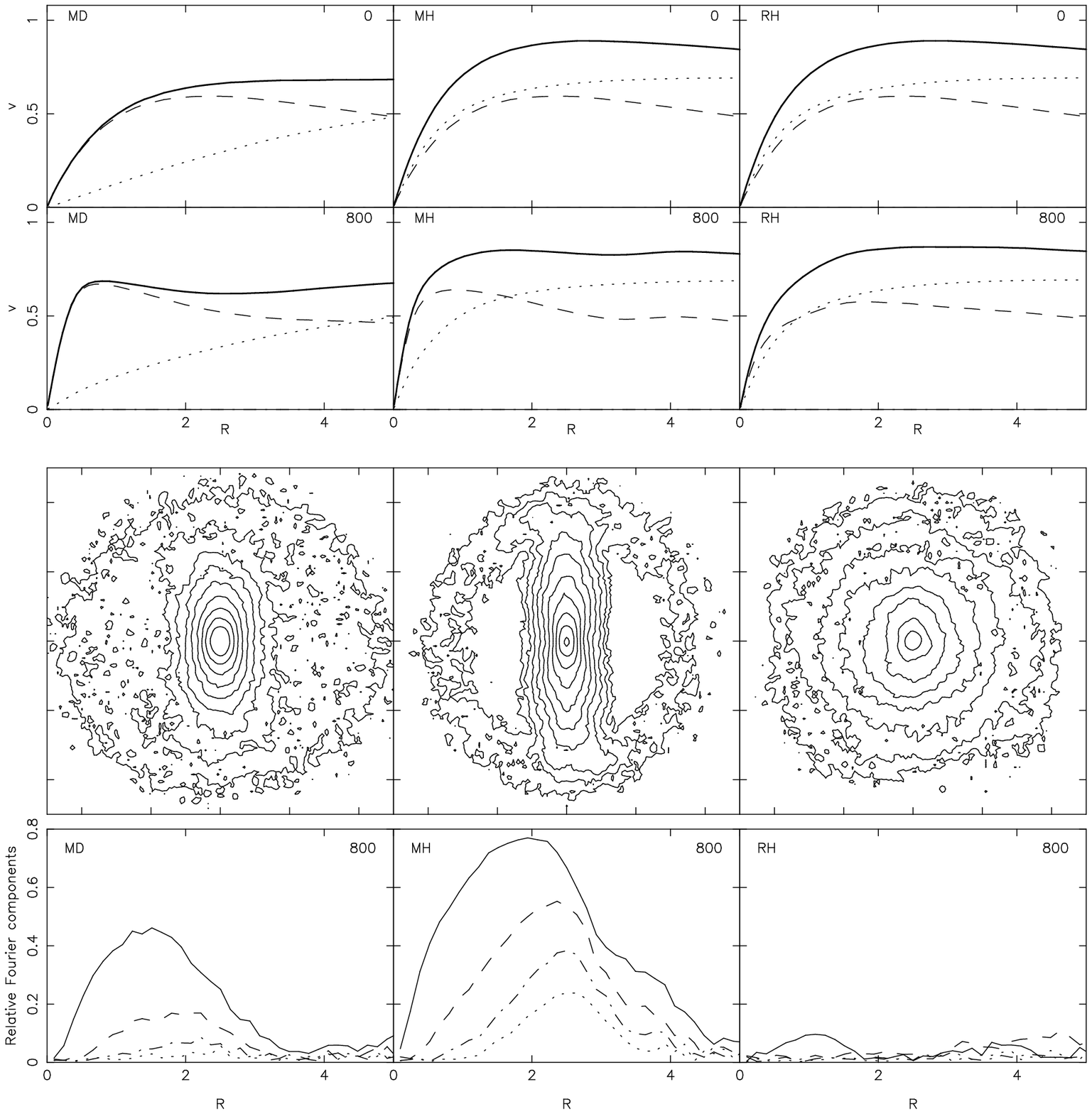}
\caption{Basic results for the three simulations discussed
  in the text. The left
    panels correspond to model MD, the middle ones to model MH and the
    right ones to model RH. The first and second row of panels give
    the circular velocity curves at $t$ = 0 and 800, respectively
    (solid lines). The contribution of the disc component is given by
    a dashed line, and that of the halo by a dotted line. 
    The third row gives, again for $t$ = 800, 
    the isodensities of the disc component when seen
    face-on. The size of the square box 
    for the third row of panels is 10 initial disc scale 
    lengths. The last row gives the relative amplitude of the $m$ = 2
(solid line), $m$ = 4 (dashed line), 6 (dot-dashed line) and 8 (dotted
line) components of the mass, again at $t$ = 800.
}
\label{fig:basic}
\end{figure}

\clearpage 

\begin{figure} 
\plotone{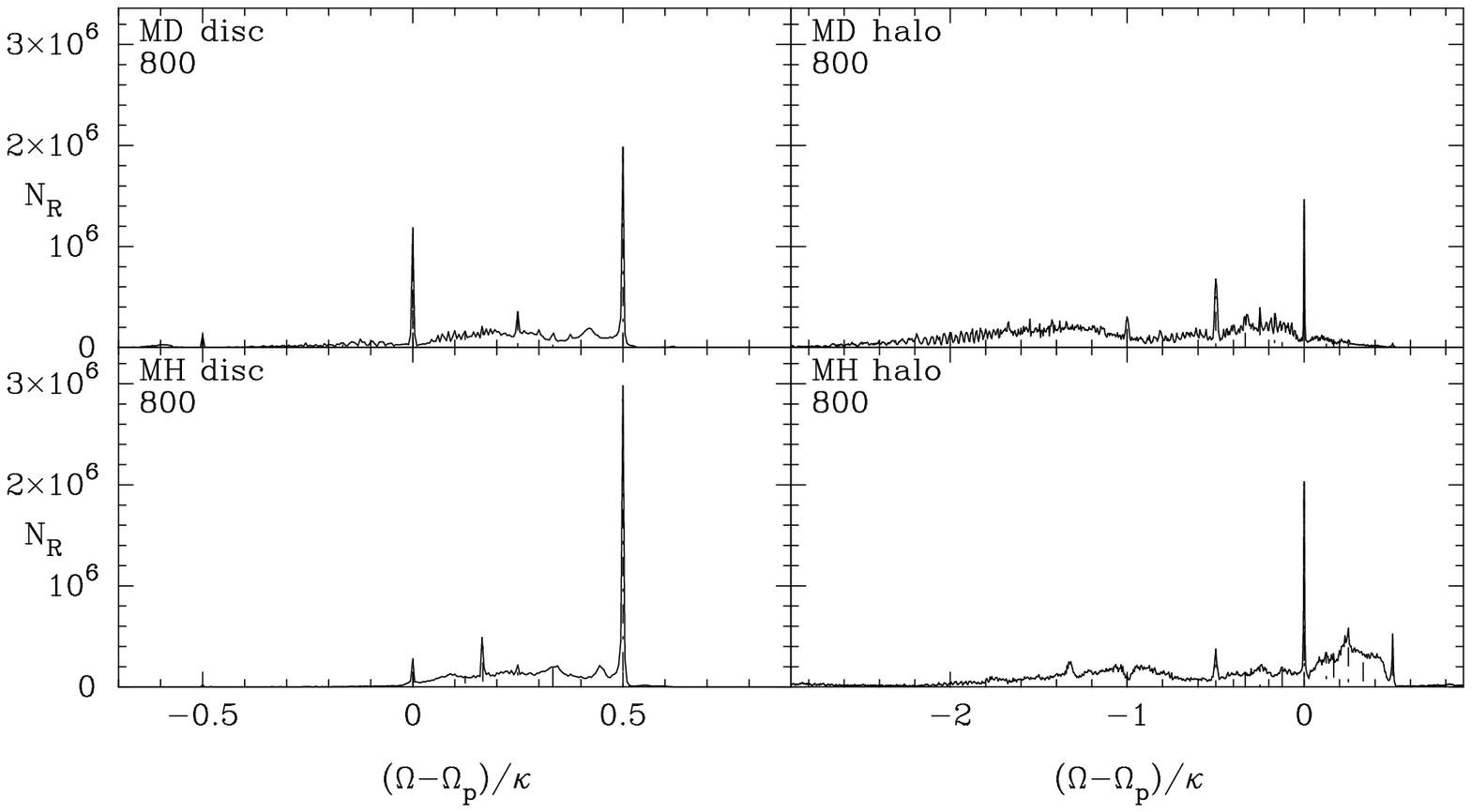}
\figcaption{Number density, $N_R$,  of particles as a
  function of the frequency ratio $R = (\Omega -
\Omega_p) / \kappa$, for simulation MD (upper panels) and MH (lower
panels) at $t$ = 800. The left panels correspond to the disc component
  and the right ones to the halo. The dot-dashed vertical lines give
  the positions of the main resonances. The results for the halo
  component have been rescaled, so as to take into account the different
  number of particles in the disc and halo components, and thus allow
  immediate comparisons.
}
\label{fig:res}
\end{figure}


\begin{thebibliography}{}
\bibitem[Athanassoula (1984)]{atha84}
  Athanassoula, E. 1984, Physics Rep., 114, 319
\bibitem[Athanassoula (1996)]{atha96}
  Athanassoula, E. 1996, in IAU Colloq. 157, Barred Galaxies,
  ed. R. Buta, D. A. Crocker \& B. G. Elmegreen (San Francisco ASP), 309
\bibitem[Athanassoula (2002a)]{atha02a}
  Athanassoula, E. 2002a, in IAU Symp. 208, Astrophysical
  Supercomputing Using Particle Simulations, eds. J. Makino \& P. Hut,
  PASP, in press, and astro-ph/0112076
\bibitem[Athanassoula 2002b] {atha02b} 
  Athanassoula, E. 2002b, in ASP Conf. Ser., The dynamics, structure
  and history of galaxies, eds. G. S. Da Costa \& E. M. Sadler,
  in press and astro-ph/0112077
\bibitem [Athanassoula \& Misiriotis 2002]{am02}
  Athanassoula, E., \& Misiriotis, A. 2002, \mnras, 330, 35 (AM)
\bibitem [Athanassoula \& Sellwood 1986]{as86}
  Athanassoula, E., \& Sellwood, J. A. 1986, \mnras, 221, 213
\bibitem [Binney \& Spergel 1982]{bs}
  Binney, J., \& Spergel, D. 1982, \apj, 252, 308 
\bibitem [Combes et al 1990]{cdfp}
  Combes, F., Debbasch, F., Friedli, D., \& Pfenniger, D. 1990, \aap,
  233, 82
\bibitem [Contopoulos \& Grosb{\o}l 1989] {cg89}
 Contopoulos G., Grosb{\o}l P., 1989, \aapr, 1,261
\bibitem [Debattista \& Sellwood 1998]{ds98}
  Debattista V. P., \& Sellwood, J. A. 1998, \apj, 493, L5
\bibitem [Debattista \& Sellwood 2000]{ds00}
  Debattista V. P., \& Sellwood, J. A. 2000, \apj, 543, 704
\bibitem [Hernquist]{hern93}
  Hernquist L., 1993, \apjs, 86, 389
\bibitem [Hernquist \& Weinberg]{hw92}
  Hernquist L., \& Weinberg. M. D. 1992, \apj, 400, 80
\bibitem [Laskar 1990]{lask}
  Laskar, J. 1990, Icarus, 88, 266
\bibitem [Little \& Carlberg]{lc}
Little, B. \& Carlberg, R. G. 1991, \mnras, 250, 161
\bibitem [Lynden-Bell \& Kalnajs 1972]{lk72}
  Lynden-Bell, D., \& Kalnajs, A. J. 1972, \mnras, 157, 1
\bibitem [Ostriker \& Peebles 1973]{op73}
  Ostriker, J. P., \& Peebles, P. J. E. 1973, \apj, 186, 467
\bibitem [Sellwood \& Wilkinson 1993]{SW}
  Sellwood, J. A. \& Wilkinson, A. 1993, Rep. Prog. Phys., 56, 173 
\bibitem [Tremaine \& Weinberg 1984]{tw84}
 Tremaine, S., \& Weinberg, M. D. 1984, \mnras, 209, 729
\bibitem [Weinberg 1985]{w85}
 Weinberg, M. D. 1985, \mnras, 213, 451
\end{thebibliography}
\end{document}